\documentclass{emulateapj}

\usepackage{graphicx} 
\usepackage{color} 
\newcommand{\be}{  \begin{eqnarray} }
\newcommand{\ee}{  \end{eqnarray} }

\def\spose#1{\hbox to 0pt{#1\hss}}
\def\lta{\mathrel{\spose{\lower 3pt\hbox{$\mathchar"218$}}
     \raise 2.0pt\hbox{$\mathchar"13C$}}}
\def\gta{\mathrel{\spose{\lower 3pt\hbox{$\mathchar"218$}}
     \raise 2.0pt\hbox{$\mathchar"13E$}}}
\begin{document}

\shorttitle{Turbulent Comptonization and Relativistic Accretion}
\title{Relativistic Accretion Mediated by Turbulent Comptonization}
\author{Aristotle Socrates\altaffilmark{1,2}}
\altaffiltext{1}{Department of Astrophysical Sciences, Princeton 
University, Peyton Hall-Ivy Lane, Princeton, NJ 08544; 
socrates@astro.princeton.edu}
\altaffiltext{2}{Lyman Spitzer Jr. Fellow}
\begin{abstract}
Black hole and neutron star accretion flows display unusually 
high levels of hard coronal emission in comparison to all other 
optically thick, gravitationally bound, turbulent astrophysical 
systems.  Since these flows sit in deep relativistic 
gravitational potentials, their random bulk motions approach 
the speed of light, therefore allowing turbulent Comptonization to 
be an important effect.  We show that the inevitable production 
of hard X-ray photons results from turbulent Comptonization in the 
limit where the turbulence is trans-sonic and the accretion power
approaches the Eddington Limit.  In this 
regime, the turbulent Compton {\it y--}parameter approaches
unity and the turbulent Compton temperature is a significant 
fraction of the electron rest mass energy, in agreement with 
the observed phenomena.

\end{abstract}

\section{General Issue and Motivation}

Like the envelopes of late-type dwarfs,
accretion disks are thought to be turbulent and optically thick.
Turbulence in stellar envelopes is driven by convection, while in
accretion disks turbulence is commonly thought to be driven by the
magnetorotational instability (MRI; Balbus \& Hawley 1991).

Accretion disks found in proto-stellar and proto-planetary systems,
cataclysmic variables, X-ray binaries and quasars/AGN all display
coronal behavior as do late-type dwarfs.  By ``corona,'' we mean the
{\it appearance}, in a spectral energy distribution, of a relatively
energetic population of particles that persistently emits radiation at
temperatures in excess of the given object's thermal photosphere.

In the case of black hole and neutron star accretion flows, the ratio
of coronal to bolometric power $L_{_{\rm c}}/L_{_{\rm bol}} \sim
0.1-1$.  In other systems that display radiatively efficient
accretion, the value of $L_{_{\rm c}}/L_{_{\rm bol}}$ is smaller by
orders of magnitude.  For example, turbulent stellar envelopes acheive
$L_{_{\rm c}}/L_{_{\rm bol}}\lesssim 10^{-3}$ in the most extreme (rapidly
rotating) case.

Here, we ask why black hole and neutron star accretion flows display
such abnormally high levels of coronal power, observed in the
form of hard power-law X-ray photons.

In the directly observable case of the solar corona, mechanical energy
leaks out through the photosphere in some combination of waves and
relatively steady magnetic structures.  
The two reservoirs of stellar mechanical energy are convective
turbulence and differential rotation.  

Since proto-planetary and white dwarf accretion disks as
well as rapidly rotating late-type dwarfs display relatively 
modest levels of coronal power, differential rotation cannot be the 
sole key ingredient in explaining the extreme coronal phenomena seen 
in relativistic accretion flows.  Furthermore, the difference between 
MRI-driven and convective turbulence must be ruled out as well 
because the non-relativistic accretion flows display only modest 
coronal phenomena.

An obvious feature that is unique to black hole and neutron star
accretion flows is that close to the central sources of 
gravity, where most of the accretion power is generated, the 
bulk motions approach relativistic values.  If the electrons
possess bulk motions that are trans-relativistic, then they 
may readily amplify soft photons up to high energies, potentially 
forming a 
hard power-law X-ray spectra though the process of 
second order bulk Comptonization.\footnote{As an aside, Cosmologists refer
to this radiative mechanism as the ``second-order kinetic 
S-Z effect.''} 

In the brief discussion that follows, 
we focus solely on the importance of turbulent
Comptonization in relativistic accretion flows.  Our approach differs
from that of Socrates et al. (2004) and Thompson (2006) in that
we completely ignore the relationship between the 
region from which soft thermal seed photons
originate -- presumeably a cool thermal disk -- with 
the region of Comptonization.  We start with a description 
of an isolated eddy undergoing turbulent Comptonization.

\section{The basic unit: an Eddy in a box}

First, we summarize the properties of an eddy in a box of 
dimension $L$.  We assume that the eddy is embedded 
in an isotropic turbulent flow.  The scale of the 
eddy $\lambda\lesssim L$ is of order the box size 
and the characteristic eddy velocity $v_{\lambda}
\lesssim c$ is taken to be trans-relativistic.  We assume that the box is 
fully ionized with a constant mass density $\rho$.  

The eddy is stirred on the outer scale $\lambda\sim L$ and 
in the absence of any form of microphysical dissipation, the 
energy injected into the scale $\lambda$ flows down to
smaller scales via a self-similar turbulent cascade.  At the
dissipation scale, the cascade cuts off and
the turbulent power is converted into heat.   
 
We now discuss some of the relevant properties of 
our eddy.

\subsection{Some Important Properties of the Eddy}

The total kinetic energy ${\mathcal K}_{\lambda}$ 
in the box at any given moment is given by 
\be
{\mathcal K}_{\lambda}\sim M_{\lambda}v^2_{\lambda}\sim 
\rho\,\lambda^3v^2_{\lambda}.
\ee
The total eddy power, or luminosity, 
$L_{\lambda}\sim {\mathcal K}/t_{\lambda}$ where 
\be
t_{\lambda}\sim
\lambda/v_{\lambda}
\ee
is the eddy turnover time.  We write the 
eddy luminosity as
\be
L_{\lambda}\sim \rho\,v^3_{\lambda}\lambda^2
\ee      
which is just the turbulent energy flux, multiplied by the 
eddy cross section.  Also, we assume that in the absence
of dissipation, the eddy velocity amplitude scales 
as a power law i.e., 
\be
v_{\lambda}\sim v_{0}\left(\lambda/\lambda_0\right)^n.
\ee
For isotropic incompressible Kolmogorov turbulence, 
the the index $n=1/3$.  We take $v_0\lesssim$ the 
speed of light $c$ and $\lambda_0\sim L$.  

Since the velocity 
amplitude $v_{\lambda}$ of our eddy is 
trans-relativistic, the inverse-Compton effect
on electrons may become important on timescales of interest, 
particularly when the 
eddy is optically thin.  Now, we assess the importance 
of the various microphysical processes arising from 
turbulent Comptonization.

\subsection{Eddy Comptonization:  Cooling Rate and  
Compton Power}

The eddy cooling rate is the rate at which photons 
remove energy from the turbulent electrons, which 
are ``stuck'' to the ions as a result of 
-- the assumed -- tight Coulomb coupling.  Therefore, 
inverse-Comptonization off of turbulent eddies may be viewed
as a microscopic dissipation mechanism for the turbulence.
Note however, that this dissipation does not necessarily occur
on scales smaller than $\lambda\sim L$.  If the eddy optical 
depth $\tau_{\lambda}\lesssim 1$ i.e., 
\be
\tau_{\lambda}\equiv \kappa_{\rm es}\rho\lambda\lesssim 1
\ee
then the photons sample the turbulent velocities on the outer
scale of the turbulent eddy, which is $\sim v_{\lambda}\sim v_0$.
Thus, the act of electrons, and thus the eddy, Compton 
up-scattering relatively cool photons necessarily 
implies that turbulent Comptonization off an eddy is 
a damping mechanism.    
 
Comptonization preserves the number of quanta, rather than the 
energy per quanta.  If the radiation field is close to isotropic, 
the Kompaneets (1957) equation, which expresses conservation
of photon number, reads
\be
\frac{\partial n_{\nu}}{\partial t}\simeq 
\sigma_{\rm es}\,n_e\,c\left(\frac{v^2_{\lambda}}{3\,c^2}
+\frac{k_BT_e}{m_ec^2}\right)\left(4\nu
\frac{\partial n_{\nu}}{\partial\nu} +
\nu^2\frac{\partial^2\,n_{\nu}}{\partial^2\nu}\right),
\label{e: kompaneets}
\ee
where $n_{\nu}$ is the photon occupation number.  The above 
expression is accurate as long as $v^2_{\lambda}\gg k_{B}T_e/m_e$
and $v^2_{\lambda}\gg E_{\nu}/m_e=h\nu/m_e$.  In other words, 
turbulent Comptonization is important as long as the contribution 
to the electron kinetic energy resulting from turbulence exceeds
the contribution from random microscopic (e.g., thermal) motions
as well as the average ``seed'' photon energy.  

Immediately, we can read off a time scale from 
eq. (\ref{e: kompaneets})
\be
t^{-1}_{\rm c,\lambda}\simeq \frac{4}{3}\sigma_{\rm es}n_e\,c
\frac{v^2_{\lambda}}{c^2},
\ee  
where $t_{c,\lambda}$ is the time it takes for the eddy to boost the
energy density of the photon field by a factor of two.  
It follows that the 
turbulent Compton power of a given eddy is simply
\be
L_{\rm c,\lambda} & \simeq & \frac{U_{\gamma}\Delta V}{t_{\rm 
c,\lambda}}
\simeq \frac{U_{\gamma}\lambda^3}{t_{\rm c,\lambda}}\nonumber\\
L_{\rm c,\lambda} & \simeq &\frac{4}{3}\sigma_{\rm es} n_e\,
c\frac{v^2_{\lambda}}{c^2}U_{\gamma}\lambda^3.
\ee
If at any point along the cascade, the turbulent Compton
power $L_{\rm c,\lambda}$ approaches the turbulent eddy power
$L_{\lambda}$, then turbulent Comptonization off of an eddy 
may be viewed as a viable microphysical dissipation mechanism
for the turbulence.

\subsection{Electron Heating Rate}

There are really three particle distributions
of concern in our box: the photons and the
bulk turbulent and microscopic thermal degrees
of freedom for the electrons.  So far, we have quantified
the rate at which energy is transferred from the
turbulence to the Comptonized radiation field.  If the
electrons are ``cold'' such that their thermal 
motions are small in comparison to their bulk motions, 
then the electrons stand a chance of heating 
up.  

In a frame moving with velocity $v_{\lambda}$,
co-moving with a given portion of an eddy, 
the thermal energy density of the gas 
$U_{\rm g}$ evolves as\footnote{Throughout, we assume that the 
electron and proton temperature are equal to one another.} 
\be
\frac{\partial U_{\rm g}}{\partial t}\simeq\sigma_{\rm es}
n_e\,c\left(\frac{\bar{E_{\gamma}}}{m_ec^2}
-\frac{4k_BT_e}{m_ec^2}\right) U_{\gamma}. 
\label{e: gas_heat}
\ee
In the above expression, $\bar{E}_{\gamma}$ is 
mean photon energy, averaged over the
photon energy spectrum.  The above expression 
conveys the action of thermal Comptonization.\footnote{One of the
most famous examples of thermal Comptonization maintaining
thermal equilibrium between radiation and gaseous matter
is in the cosmic background radiation, before and during the 
moments of recombination.}  

We are 
interested in the regime where turbulent Comptonization
boosts relatively few photons up to 
high energies, where $E_{\gamma}\lesssim 
m_ec^2$.  Though few in number, their large 
energies allow them to carry away a significant -- 
if not dominant -- fraction of the photon 
power.  If the gas is cold such that the bulk
-- and therefore, turbulent -- motions are large in 
comparison to the particle's thermal 
motions, then $\bar{E}_{\gamma}\gg k_BT_e$.  
In that case, we identify a Compton heating rate
\be
t^{-1}_{\rm c,heat}\simeq \sigma_{\rm es}\,n_e\,
c\frac{\bar{E}_{\gamma}}{m_ec^2}\frac{U_{\gamma
}}{U_{\rm g}},
\label{e: t_heat}
\ee      
which is the rate at which energetic 
Comptonized photons heat up up the 
gas, increasing its thermal content per 
unit volume.  

\section{An Eddy around a black hole}

Now, we place our eddy near a black hole
to get a feel as to how an ensemble of 
eddies -- that collectively form an 
accretion flow -- behaves.  We
place our eddy at a radius $R=R_G/\epsilon$
from the hole, where $R_G=GM_{\bullet}/c^2$ and
$\epsilon$ is the gravitational radius and 
radiative efficiency of the hole, respectively.
The radiative efficiency $\epsilon$ 
is roughly determined by 
the location of inner-most stable circular orbit
$R_{isco}$ i.e., $\epsilon\sim R_g/R_{isco}$.  So, 
if the $\epsilon\sim 0.1$ then roughly $\sim 100$ MeV
worth of gravitational binding energy per baryon 
is available for conversion into other forms
of energy.  The discussion that follows applies to 
neutron star accretion flows as well.  Since black holes
do not have a surface, the overall physical system is
simpler and for this reason, we restrict our 
discussion to black hole accretion flows.  

\subsection{Accretion power and accretion timescale}

We imagine that our eddy is surrounded by other eddies of 
comparable size and shape.  The interaction between 
eddies not only contributes to the turbulent
cascade of energy to small scales, but also
transfers angular momentum between 
radially adjacent eddies. 

The transfer of angular momentum, the viscous 
generation of accretion power and the release of 
gravitational binding energy are connected by 
the expressions that govern angular momentum and 
energy balance.  The viscous dissipation rate per 
unit volume $\varepsilon^+$ resulting from 
an effective turbulent viscosity is given by
\be
\varepsilon^+\simeq \rho\,v^2_{\lambda}\frac{{\rm d}\Omega}{
{\rm d}{\rm ln}R}\sim \rho\,v^2_{\lambda}\Omega,
\ee
which is just the kinetic energy density,
multiplied by the shear rate.  Note that in 
the above expression, we assume that each respective
eddy is rotating at the Keplerian rate, with an
angular velocity $\Omega\simeq\sqrt{GM_{\bullet}/R^3}$.  
The total accretion power, or accretion luminosity
$L_{\rm acc,\lambda}$ for the eddy is given by
\be
L_{\rm acc,\lambda}\simeq \rho\,\lambda^3v^2_{\lambda}
\,\Omega\simeq {\mathcal K}_{\lambda}\Omega.
\label{e: acc_power}
\ee 

Of course, each eddy is a transient structure 
embedded in a continuum of matter that slowly 
drifts towards the hole.  This drift, which 
results from the transport of angular momentum
is quantified by the specific 
angular momentum $\ell$ and the
mass accretion rate through the eddy, $\dot{M}_{\lambda}$,
in the following way
\be
\dot{M}_{\lambda}\,\ell\simeq \lambda^3\,\rho\,
\,v^2_{\lambda},
\label{e: ang_transf}
\ee
where we have ignored the 
contribution from the inner boundary.  As an interesting
aside, we see that $\dot{M}_{\lambda}\ell\simeq
{\mathcal K}_{\lambda}$ or, the rate of angular 
momentum loss within an eddy is equal to 
the kinetic energy, ${\mathcal K}_{\lambda}$,
of the eddy itself.
The combination of the expression 
for viscous dissipation, eq. (\ref{e: acc_power}),
and angular momentum conservation, 
eq. (\ref{e: ang_transf}), allow
us to connect the accretion power
$L_{\rm acc,\lambda}$ to the accretion 
rate $\dot{M}_{\lambda}$ through the eddy i.e., 
\be
L_{\rm acc,\lambda}\simeq {\mathcal K}_{\lambda}\Omega\simeq
\dot{M}_{\lambda}\,\ell\,\Omega\simeq \frac{G\dot{M}_{\lambda}
M_{\bullet}}{R},
%\simeq \epsilon\dot{M}_{\lambda}\,c^2.  
\ee
analogous to the well known result.

The characteristic time for an eddy to deplete
its mass $t_{\rm acc,\lambda}$ is obtained by 
re-writing conservation of angular
momentum, eq. (\ref{e: ang_transf}), as
\be
\dot{M}_{\lambda}\,\ell\simeq \frac{M_{\lambda}}{
t_{\rm acc,\lambda}}\, R^2\,\Omega
 & \simeq & M_{\lambda}\,v^2_{\lambda}
\simeq {\mathcal K}_{\lambda}\nonumber\\  
t^{-1}_{\rm acc,\lambda} &  \simeq & 
\frac{v^2_{\lambda}}{R^2\Omega},
\label{e: t_acc}
\ee  
which is often written in terms of 
an anomalous viscosity and a 
Shakura \& Sunyaev (1973) $\alpha$-parameter.
  
\subsection {An Eddington-Limited Eddy}

Now, we determine under what conditions the 
various dynamical, thermal and radiative
timescales become competitive with one another.
Throughout, we specify that the turbulent 
motions on the stirring scale $\lambda$ are
trans-relativistic -- a necessary condition 
for turbulent Comptonization to be a powerful
effect.  Another condition, is that the optical 
depth must at least be of order unity, in which 
case of the rate of photon escape is equal to the 
rate of energy gain, yielding a relatively 
flat (in $\nu\,F_{\nu}$) and powerful 
Comptonized spectrum.  

We take the limit where $\lambda\sim R\sim R_g/
\epsilon$ and $v^2_{\lambda}\lesssim c^2_s\sim 
\epsilon \,c^2$.  That is, we approximate that the 
eddy is in dynamic balance with the black hole's 
gravitational field at $R/R_g\sim 
\epsilon^{-1}$ gravitational radii from the 
system's gravitational center.    
In order to 
calculate the eddy optical depth, or 
alternatively the mean density, we further 
assume that the eddy is radiating at the 
Eddington limit for the eddy $L_{\rm edd,\lambda}$ i.e., 
\be
L_{\rm acc,\lambda}\simeq \frac{G\dot{M}_{\lambda}M_{\bullet}}{
R}\simeq\epsilon\,\dot{M}_{\lambda}\,c^2\simeq 
L_{\rm edd,\lambda}\simeq \frac{G\,M_{\bullet}\,c}
{\kappa_{\rm es}}.
\ee  
Note that $L_{\rm edd,\lambda}$ differs from the 
Eddington limit $L_{\rm edd}$ for the entire 
flow by a factor of $1/4\pi$ since the eddy on 
subtends $\sim 1$ steradian of the solid angle 
around the black hole.  Putting together the
expression above with eq. (\ref{e: t_acc}) we may write
the eddy optical depth as
\be
\tau_{\lambda}\sim \epsilon^{-1/2}\gtrsim 1.
\label{e: tau}
\ee  
At the Eddington limit, the Thomson 
optical depth across the eddy $\tau_{\lambda}$ 
is of order unity and therefore, the 
photon field ``samples'' the energy bearing 
eddies on the outer scale $\lambda$.  Therefore, the 
rate for turbulent Compton heating $t^{-1}_{\rm c,
\lambda}$ and the turbulent Compton 
luminosity $L_{\rm c,\lambda}$ should 
be evaluated by using the outer scale 
length $\lambda $ and the 
outer scale velocity $v_{\lambda}$.  Finally, for an 
Eddington-limited eddy, we find
\be
t^{-1}_{\lambda}\simeq t^{-1}_{\rm c,\lambda}
\simeq t^{-1}_{\rm acc,\lambda}\simeq
t^{-1}_{\rm dyn},
\label{e: edd_eddy1}
\ee   
where $t_{\rm dyn}\sim c_s/L\sim 1/\Omega$.
Furthermore,
\be
L_{\rm c,\lambda}\simeq L_{\rm acc,\lambda}
\label{e: edd_eddy2}
\ee  
for an Eddington-limited eddy.

\begin{figure}
\begin{center}
\includegraphics{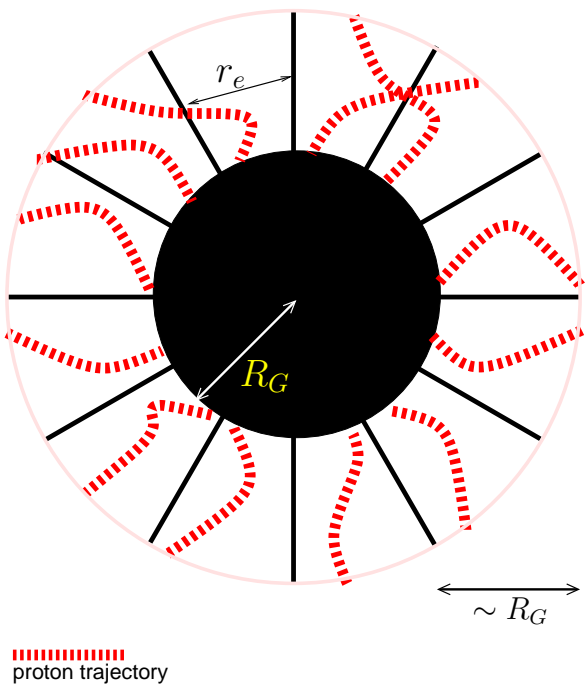}
\caption{Cross section of critical accretion at the Eddington 
Limit $L_{_{\rm Edd}}$onto a black 
hole with radiative efficiency $\epsilon=1$.  
The flow is divided into ``Thomson units'' in order to depict the
fact that the number of protons $N_p$ responsible for fueling 
an Eddington's worth of power in one light crossing time 
$\sim R_{G}/c$ is given by the ratio of the area of 
the black hole to the Thomson cross section.  That is, 
$N_p=N_e\sim R^2_{G}/r^2_e$ where $r_e=e^2/m_ec^2$ is the 
classical electron radius.  In order to 
maintain a high radiative
efficiency, $\epsilon
\sim 1$, the orbit of each individual proton must undergo 
a deflection of order $\sim$ one $R_G$, which implies that 
$\Delta v^2/c^2\sim v^2_{\lambda}/c\sim 1$.  Together, this 
implies that the turbulent Compton y--parameter 
$y_{\lambda}\sim 1$, leading to an observed 
flat spectral index $\Gamma\sim 2$.}
\label{f: y_turb}
\end{center}
\end{figure}

The two expression, eqs. (\ref{e: edd_eddy1})
and (\ref{e: edd_eddy2}) above, are interesting results.  Together, 
they indicate the release of gravitational energy 
resulting from turbulent viscous dissipation is 
equally matched by the Compton power that is 
generated by the turbulence as well.  Therefore, it
is possible that all of the fundamental physical 
mechanisms that mediate accretion i.e., the transport
of angular momentum, randomization of gravitational 
binding energy and the conversion of that energy into 
photon power is entirely determined by a single 
physical phenomena: turbulence.

However, our picture is not complete.  Bathing the eddy in 
such an intense (self-generated) hard radiation 
field has consequences for the thermal motions of the 
gas.  Consider the form of the Compton heating
rate, given by eq. (\ref{e: t_heat}).  The act of turbulent 
Comptonization boosts the mean photon 
energy to a value $\bar{E}_{\gamma}\lesssim 
m_ev^2_{\lambda}$, which implies that 
\be
t^{-1}_{\rm c,heat}\simeq\frac{U_{\gamma}}{U_g}\,
 t^{-1}_{c,\lambda}\gg t^{-1}_{\rm c, \lambda}
\ee
for an Eddington-limited eddy since $U_{\gamma}>>U_{g}$
even if $k_BT_g\sim \bar{E}_{\gamma}$.  This results from 
the fact that 
the radiation pressure is roughly a factor 
$m_p\,c^2/k_BT_p$ larger than the
gas pressure, at the Eddington limit.

The large value of the heating rate $t^{-1}_{\rm c, heat}$
leads to the conclusion that in equilibrium, the thermal
energy per baryon equals the average energy of the hard 
Comptonized photons.  In that case, the source term on 
the right hand side of eq. (\ref{e: gas_heat}) approaches
zero and thermal equilibrium for the gas is reached.

Altogether, the overall picture for an Eddington-limited 
eddy is different from the viewpoint presented in 
Socrates et al. (2004).  Apparently, it seems unlikely 
that gaseous species can persist at specific thermal 
energies that are significantly smaller than 
$v^2_{\lambda}$.  Therefore, turbulent Comptonization 
in black hole and neutron star accretion flows may also 
be viewed as an electron heating mechanism as well.

%\section{Thermal energy of electrons and protons}

\section{The turbulent Compton 
y-parameter and the meaning of its value
near a black hole}
 
We continue our assessment of the Eddington-limited
eddy near a black hole 
introduced in the last section.  The spectral 
features of saturated and mildly unsaturated 
Comptonization are well described by the slope of 
a self-similar Comptonized spectral feature and 
a high energy cutoff.  The slope is often expressed in 
terms of the Compton $y$-parameter, given by the fractional 
energy shift per scattering multiplied, by the average number
of scatterings.  When $y\sim 1$, the ``input'' spectrum, 
responsible for providing the soft seeds, is deformed by the 
addition of a hard self-similar power-law component.  
In such an event, the ratio of hard Comptonized to soft seed
power is also of order unity.  When the eddies 
are optically thin 
such that $\tau_{\lambda}\leq 1$, we define the 
turbulent Compton $y-$parameter as
\be
y_{_\lambda}=\frac{4}{3}\frac{v^2_{\lambda}}{c^2}\,\tau_{\lambda}
\simeq \frac{4}{3}\,\epsilon^{1/2}\lesssim 1.
\ee
The important result above, that the turbulent Compton 
$y-$parameter is close to unity, is not an accident.  

Consider the simple representation
of extreme quasi-spherical black hole 
accretion given by Figure \ref{f: y_turb}.
For illustrative purposes, the hypothetical case where
accretion is $100\%$ efficient i.e., $\epsilon =1$, is
depicted.  Thus, dynamical effects arising from general 
relativity, such as the presence of
the last stable circular orbit, are completely ignored. 
Critical accretion, which implies $t^{-1}_{{\rm acc,\lambda}}
\sim t^{-1}_{\rm dyn}$, while
unit efficiency, $\epsilon =1$, implies that 
$\lambda\sim R\sim R_{G}$.  
For this case, the number of protons 
$N_p$ responsible for fueling an Eddington's worth of accretion 
power is roughly given by $N_p\sim A_{\bullet}/\sigma_T
\sim R^2_{G}/r^2_e$, where $A_{\bullet}$ and $r_e$ are 
the surface area of the black hole and the classical 
electron radius, respectively. 

It follows that Comptonized photons remove the accretion 
power of each individual proton on average scatter once
with an electron -- as long as the number of electrons
equals the number of the protons.  Or in other words, 
$\tau_{\lambda}\sim 1$ for highly efficient critical 
accretion.  At the same time, the value of randomized proton 
velocity squared $\Delta v^2\sim v^2_{\lambda}\sim c^2$, 
since of order $m_p\,c^2$ worth of binding energy is removed
per proton.  The combination of these two attributes 
of critical efficient accretion onto a black hole yields the 
result that $y_{\lambda}\sim 1$.  Hence, {\it 
Eddington-limited relativistic accretion 
is a natural turbulent Compton amplifier}.

%\section{Thermal Energy of Electrons and 
%Protons}

%Now, it is clear that turbulene in the vicinity of 
%a black hole or neutron star may not only transfer 
%angular momentum but may also be directly 
%responsible for the converstion of gravitational 
%into photon power.  Nevertheless, the result th  

\section{Some Open questions and Conclusions}

At radii immediately 
further out from the inner-most stable circular orbit, 
the ability for trans-sonic turbulence to Comptonize soft 
photons diminishes since $v^2_{\lambda}$ decreases, while 
$\tau_{\lambda}$ increases.  The latter implies that the input 
photons can only ``sample'' eddies of lower energy, on smaller
scales (Socrates et al. 2004).  
Also at larger radii, the ability for the flow 
to absorb a photon via free-free and bound-free processes 
increases (Socrates et al. 2006).  
We suspect that these adjacent outer regions serve 
as the source of -- presumably thermal --soft 
input photons for the centrally located Comptonizing eddies.
In the case accretion onto a neutron star, thermal emission 
from the stellar surface is additional source of soft seeds. 

We have yet to specify the actual source
of trans-sonic turbulence itself.  In the familiar language of 
the Shakura \& Sunyaev (1973) thermal disk model, we have
assumed throughout that $\alpha\lesssim 1$ i.e., 
$v^2_{\lambda}\lesssim c^2_s$.  It is now 
widely accepted that the MRI 
(Balbus \& Hawley 1991) generically leads to a turbulent 
accretion stress, allowing for the transfer
of angular momentum to occur 
in highly conducting fluids.  However, simulations 
of the MRI typically find low values for the
Shakura \& Sunyaev (1973) $\alpha$-parameter 
such that $\alpha \sim 0.1-0.01$, though a clear 
understanding for the value of the MRI's $\alpha$ 
is not currently in hand.   
  
Our main results for Eddington-limited trans-sonic turbulence near a
relativistic source of gravity are as follows:

1) Turbulence is responsible
for transporting angular momentum 
while simultaneously converting gravitational 
power of protons into Comptonized photon power.

2) The dynamical, accretion, Compton cooling and 
eddy overturn time are all comparable to one another.

3) The turbulent Compton power 
of each individual eddy is comparable to the
accretion power of that eddy.

4) Electrons will inevitably heat up to the a
significant fraction of the electron virial temperature.
In other words, turbulent Comptonization is an important 
electron heating mechanism.

5) The $y-$parameter due to turbulent Comptonization
is close to unity.  Therefore, the outgoing hard 
X-ray spectrum is a power law with flat spectral 
index $\Gamma\sim 2$, in agreement with observations. 

6) The results above are independent of black hole mass.

\acknowledgements{The author thanks P. Goldreich for useful conversations
and S. Davis for suggestions on the manuscript.
Support from a Lyman Spitzer Jr. Fellowship 
awarded by the Department of Astrophysical Sciences at Princeton 
University and a Hubble Fellowship, administered by the Space Telescope 
Science Institute is appreciated.  Finally, the author thanks the 
Institute for Advanced Study in Princeton, NJ 
for its hospitality as a portion of this work was completed there.}

\end{document}